\documentclass[aps,prl,twocolumn,showpacs,10pt,superscriptaddress,preprintnumbers,nofootinbib]{revtex4-1}
\usepackage{graphicx}% Include figure files
\usepackage{dcolumn}% Align table columns on decimal point
\usepackage{bm}% bold math
\usepackage{fancyhdr}
\usepackage{amsmath,amsfonts,extarrows,amssymb,mathrsfs,times,bm}
\usepackage{extarrows}
\usepackage{graphicx}
\usepackage{float}
\usepackage{graphicx,epstopdf}% Include figure files
\usepackage{dcolumn}% Align table columns on decimal point
\usepackage{exscale}
\usepackage{relsize}
\usepackage{mathtools}
\usepackage{mhchem}
\usepackage[colorlinks=true,breaklinks=true,linkcolor=blue,citecolor=blue,urlcolor=blue]{hyperref}

\pdfoutput=1

\usepackage[export]{adjustbox}
\usepackage{xcolor}
\usepackage{graphicx}% Include figure files
\usepackage{dcolumn}% Align table columns on decimal point
\usepackage{bm}% bold math
\begin{document}
\title{Affine gravitational scenario for dark matter decay}% Force line breaks with \\
\author{Hemza Azri}
\email{hmazri@uaeu.ac.ae}%Lines break automatically or can be forced with \\
\affiliation{Department of Physics, United Arab Emirates University,
Al Ain 15551 Abu Dhabi, UAE}%
\author{Adil Jueid}
\email{adil.hep@gmail.com}%Lines break automatically or can be forced with \\
\affiliation{Department of Physics, Konkuk University, Seoul 05029, Republic of Korea}%
\author{Canan Karahan}
\email{ckarahan@itu.edu.tr}%Lines break automatically or can be forced with \\
\affiliation{Physics Engineering Department, Istanbul Technical University, 34469 Maslak, Istanbul, Turkey}%
\author{Salah Nasri$^{1}$}
 \email{snasri@uaeu.ac.ae}
\affiliation{International Center for Theoretical Physics, Strada Costiera 11, I-34151 Trieste, Italy}%

\date{\today}% It is always \today, today,

\begin{abstract}
Since the current evidence of its existence is revealed only through its gravitational influence, the way dark matter couples to gravity must be then of primary importance. Here, unlike the standard model sector which is typically coupled to metric, dark matter is supposed to couple only to spacetime affine connection through a $Z_{2}$-symmetry breaking term. We show that this structure leads to a coupling between dark matter, which is considered scalar, and the standard model Higgs potential. This induces dark matter decays into standard model particles through the Higgs which acts as a portal between the visible and the dark sectors. We study thoroughly the resulting decay modes for various mass ranges, and provide relevant bounds on the nonminimal coupling to affine gravity in line with observational data. Moreover, we find that the coupling to Higgs can be sufficiently large to facilitate production of dark matter lighter than $10$ GeV at current and future  high energy colliders.

\end{abstract}

%\pacs{Valid PACS appear here} PACS, the Physics and Astronomy
\maketitle
\thispagestyle{fancy}
\lhead{arXiv preprint}
%\rhead{Prepared for submission}
%\tableofcontents
\emph{Introduction}\textemdash   The current standard model of cosmology suggests that the dominant mass of the observable universe, manifesting as a non-luminous matter, has an effects only through its gravitational interactions \cite{planck}. This dark matter (DM) tends to be a central component not only in explaining rotation curves in the outer of galaxy halos as it has been revealed for the first time \cite{rotation:curves}, but also in the current view of structure formation. In fact, strong observational evidence shows that clusters of galaxies might be embedded in an immense distribution of this invisible matter leading to the fact that at early times, primordial seeds are gradually gathered inside clumps of DM to finally form galaxies thanks to gravitational instability \cite{peebles}. Nonetheless, at the fundamental level, postulating the existence of a new ingredient to account for a missing mass in the observable universe requires a new physics beyond the standard model (SM). In this respect, neither the nature of DM nor its properties is sufficiently known. On the other hand, the lack of any mechanism by which DM has been produced in the very early universe suggests that its lifetime may exceed the age of the universe itself. To that end, for general scenarios new symmetry is to be imposed that guarantees DM stability and ensures a relevant long lifetime.

Models of DM are usually formulated in the framework of field theory in flat spacetime which is sufficient and relevant for studying its properties via scattering processes and decays to SM particles. Nevertheless, one may have to consider DM dynamics in curved spacetime as an essential requirement since it acts only through its gravitational effects. This argument was considered recently where it has been shown that DM could decay to SM particles, not through a \textit{prior} interactions with the latter, but only through nonminimal coupling to gravity in terms of effective operators proportional to DM field and suppressed by powers of Planck mass \cite{ibarra}. Since both SM and DM are coupled to spacetime metric, the transition to Einstein frame (metric transformation) ensures that DM decays to all SM particles with a tiny total decay rate safeguarding a long lifetime. This mechanism is not restricted to scalar DM but can be also realized for fermionic DM as long as gravity is metrical \cite{ibarra}. 

The aim of this letter is to set up a novel way for the coupling of DM to gravity and examine the possibility of its decays through nonmiminal couplings within the new picture. The standard model particles are known to be merely accommodated in curved background in a general covariant form, structured by couplings to the spacetime metric that provide a consistent interaction to gravity. In this work we will rather investigate the possibility where DM interacts only with the spacetime connection which is not related to the metric a \textit{priori}. This novel way of coupling to gravity allows one to distinguish a possible viable scenario for DM which will simply manifest as a scalar. In fact, fermions and vector fields which are known to be intimately connected to the spacetime metric in their familiar Lagrangians, can be hardly ever accommodated in purely affine spacetime. This also reflects the fact that it is only the SM-Higgs (the only scalar) that can be easily placed in this metricless background. We will show that a $Z_{2}$-breaking symmetry term which arises only through the affine curvature as a nonminimal coupling of DM, induces decays to only Higgs boson with decay rates suppressed by Planck mass. We will study the  different branching ratios of DM decays and distinguish various interesting mass ranges for its lifetime to be sufficiently longer than the age of the universe. 

\emph{Affine gravitational dark matter}\textemdash   The SM action is written in curved spacetime by imposing the equivalence principle where the flat spacetime metric is replaced by the curved metric $g$, and reads 
\begin{eqnarray}
\label{sm action}
S=\int d^{4}x
\label{sm action}
\sqrt{|g|}\,\Big\{
\frac{M^2_{Pl}}{2}g^{\mu\nu}R_{\mu\nu}(g) +L[g,\text{SM}] \Big\}
\end{eqnarray}
where the first term refers to the Einstein-Hilbert action whilst the SM Lagrangian takes the form
\begin{eqnarray}
\label{SM_lag}
L[g,\text{SM}]=
&&-g^{\mu\nu}(D_{\mu}H)^{\dagger} (D_{\nu}H) -V(H) -i\bar{\psi}g^{\mu\nu}\gamma_{\mu}\bm{\nabla}_{\nu}\psi \nonumber \\ &&-\frac{1}{4}g^{\mu\nu}g^{\alpha\beta} F^{a}_{\mu\alpha}F^{a}_{\nu\beta}. 
\end{eqnarray}

This structure is known as the minimal coupling to gravity, and it shows how the SM fields are tightly connected to the spacetime metric. Likewise, there is no general symmetry that prevents the presence of SM-curvature coupling terms \cite{demir_SM}.  

On the other hand, it is known that the only current evidence for the existence of DM is through its gravitational effects, and hence knowing the way it couples to gravity is crucial and essential by itself. In this regard, one may have to go beyond the standard form (\ref{SM_lag}) when DM is considered in the first place. In this letter, we will propose a possible extension of the SM by a Lagrangian density that represents the DM sector. First, one has to notice that the familiar structure of the SM action stands on a specific picture of gravity, the purely metric gravity, where all the geometric quantities such as the connection and curvature are given in terms of the metric. Therefore, spacetime geometry in this form is subject to strong restriction where it is considered Riemannian from the beginning. However, the geometric aspect of gravity pertaining to spacetime curvature does not impose a metrical structure \textit{a priori}. Thereby, one might rather reduce the restrictions on the spacetime geometry. In other words, the spacetime must be endowed not only with a metric tensor but with also an affine connection as an independent field. This provides us with both metric and affine parts when forming Lagrangians of various physics models.  
In the following, we simply assume that the SM sector retains its familiar coupling to metric, but we propose that DM is accommodated by the purely affine part of spacetime, which  can be a natural  way for DM to be  separate from the SM  matter and gauge sectors. 

The first step towards constructing a purely affine model of DM is to examine the possibility of decoupling it from the metric tensor. To that end, one may first place the DM field in a generic metric-affine Lagrangian density where the metric is non-dynamical so that it breaks up with DM smoothly. Here, we demonstrate that while vector and fermionic fields stay intimately attached to metric, scalars in contrast break up easily with it. To show this, one starts with the following toy model 
\begin{eqnarray}
\label{palatini}
\mathcal{L}[\mathcal{I}]=\sqrt{|g|}\,\Big\{
\frac{M^2}{2}g^{\mu\nu}R_{\mu\nu}(\Gamma)
-\frac{1}{2}g^{\mu\nu}X_{\mu\nu} -U(\mathcal{I}) \Big\}
\end{eqnarray}
where $R_{\mu\nu}$ is the Ricci curvature of the affine connection $\Gamma$ and $U(\mathcal{I})$ is the potential of dark matter field $\mathcal{I}=\phi, \chi, A$, referring to scalar, fermionic or vector dark matter respectively, whereas its kinetic term is given in familiar form by
\begin{align}
\label{kinetic parts}
X_{\mu\nu}(\mathcal{I})= 
\begin{cases}
    \nabla_{\mu}\phi\nabla_{\nu}\phi     & \text{for scalar DM } \\
    \bar{\chi}\gamma_{\mu}\bm{\nabla}_{\nu}\chi              & \text{for fermionic DM } \\ 
    \frac{1}{2}g^{\alpha\beta}\mathbb{F}_{\alpha\mu}\mathbb{F}_{\beta\nu}    & \text{for vector DM }
\end{cases}
\end{align}

It is clear that the above  Lagrangian  has the relevant form of metric-affine gravity that leads to Einstein's field equations for every DM scenario listed above. Additionally, one clearly notices the metric-dependence of the kinetic terms of the vector and fermionic DM, where in the latter it arises within the covariant derivative through the vierbein field. In contrast, the scalar DM kinetic part in (\ref{kinetic parts}) shows no dependence on the metric. Now, in purely affine spacetime, forming an action for DM requires a Lagrangian density free of metric. Therefore, the next step is to decouple the DM field from the metric with the aid of (\ref{palatini}), thus integrating out the metric via the equations of motion of the DM field. It turns out that this becomes possible only in the case of scalar dark matter,  denoted by $\phi$, where it can be easily checked that the gravitational equations arising from (\ref{palatini}) lead to  
\begin{eqnarray}
\label{metric}
g_{\mu\nu} = \frac{M^2 R_{\mu\nu}(\Gamma)-X_{\mu\nu}(\phi)}{U(\phi)}.
\end{eqnarray}

Integrating out the metric in this way ensures that scalar DM can be reincorporated into (\ref{palatini}) in a metricless form, enabling for a purely affine description. Hence, among various fields, those enjoying a complete structure in purely affine spacetime are scalars. Therefore, using (\ref{metric}), the metric now is disentangled from DM leading to
\begin{eqnarray}
\label{DM_lag}
\mathcal{L}[\Gamma,\text{DM}]=\frac{\sqrt{\left| M^{2} R_{\mu\nu}\left(\Gamma\right)- \nabla_{\mu}\phi\nabla_{\nu}\phi  \right|}}{U(\phi)}.
\end{eqnarray}
which  describes a scalar field coupled minimally to gravity without metric, and it leads to Einstein's field equations (providing $M=M_{Pl}$) where the connection tends to be (\textit{a posteriori}) the Levi-Civita of the metric (\ref{metric}). We have to emphasize here that the purely affine density (\ref{DM_lag}) can be formed using only general covariance instead of the previous derivation in the presence of metric. In this case, the latter will be generated dynamically. This metricless model of scalar fields has been examined recently in details in line with some cosmological applications \cite{affine:inflation}. Here and in the rest of the work, the connection and the Ricci curvature are taken both symmetric. It is necessary however to mention here that there have been various but different attempts for general relativistic theories of fermions and vector fields without referring to metric \cite{fermions:without:matter}. 

\emph{Dark matter decay through affine gravity portal}\textemdash  The aim of the following part is to study the decays of the scalar DM through gravity portals arising from nonminimal couplings to gravity. For that, we will assume that DM remains stable due to global symmetry which can be broken explicitly by the presence of nonminimal coupling terms. However, unlike metric gravity case (see \cite{ibarra} for details), this nonminimal interaction must come out only from the affine part  since it is the only part that contains DM. 

Nevertheless, the fact remains that these gravity portals will not induce DM decays to the SM particles if the two sectors remain fully separate. Indeed, any transformations (in terms of only dark matter) of the affine connection or curvature that arise within (\ref{DM_lag}) does not affect the SM Lagrangian. To that end, the possibility of tying-in the two sectors must be perceived in accord with the above requirements. As argued previously, the form of the affine Lagrangian supports only scalar fields, and if to be extended by the SM sector, then only scalar parts of the latter can be easily accommodated. It is clear that (\ref{DM_lag}) would lead to the same equations of motion for the Higgs boson which is the sole scalar field in the SM sector, yet the kinetic terms of the Higgs field contains gauge bosons (vector fields) that are not easily supported here. Thereby, the most relevant way where the SM is linked to DM sector in the present scenario is to incorporate the Higgs potential into the affine gravity part. Thus, the previous Lagrangian can be extended to
\begin{flalign}
\label{nonminimal coupling}
\mathcal{L}[\Gamma,\text{DM}]= \frac{\sqrt{\left| M^{2} R_{\mu\nu}\left(\Gamma\right)- \nabla_{\mu}\phi\nabla_{\nu}\phi + \xi M \phi R_{\mu\nu}\left(\Gamma\right) \right|}}{V(\phi,H)} \nonumber \\
\end{flalign}
where $\xi$ is a dimensionless constant, and  $V(\phi,H) $  is the extended potential 
\begin{eqnarray}
\label{extended potential}
V(\phi,H) \supset U(\phi)+ m^{2}_{H}H^{\dagger}H + \lambda_{H}(H^{\dagger}H)^{2}.
\end{eqnarray}

The expression (\ref{nonminimal coupling}) describes a DM nonminimally coupled to gravity through a linear term in the Ricci curvature of the affine connection that represents the simplest first order $Z_{2}$-symmetry breaking term, and improved by the Higgs potential as a relevant scalar. It is remarkable that this structure does not allow \textit{a priori} for what is known as Weyl (or conformal) transformation by which one switches between conforomal frames. Therefore, an important feature of this structure is that the transition to minimal coupling case (\ref{DM_lag}) in which the DM-curvature interaction term disappears, is realised here without applying any metric transformation. In fact, Lagrangian (\ref{nonminimal coupling}) is simply equivalent to  
\begin{flalign}
\label{rescaled lag}
\mathcal{L}[\Gamma,\text{DM}]= \frac{\sqrt{\left| M^{2} R_{\mu\nu}\left(\Gamma\right)- \bm{\alpha}^{-1}(\phi)\nabla_{\mu}\phi\nabla_{\nu}\phi  \right|}}{\bm{\alpha}^{-2}(\phi)V(\phi,H)}
\end{flalign}
where for ease of writing we have introduced a dimensionless parameter $\bm{\alpha}(\phi)= 1+\xi \phi/M$. As a result, the first order $Z_{2}$-symmetry breaking term induces DM decays to SM-Higgs through the induced interactions $\bm{\alpha}^{-2}(\phi)V(\phi,H)$ even in the absence of any explicit interaction of DM with SM particles as it is clear from  (\ref{extended potential}). The  interactions are, however, a direct effects of the nonminimal coupling to affine gravity when the latter is recast in  a canonical form. One then notices that depending on the DM mass, various decays into SM particles arise through Higgs boson.

\begin{figure}[tbp]
\centering
\includegraphics[width=9.5cm,height=9.5cm,keepaspectratio]{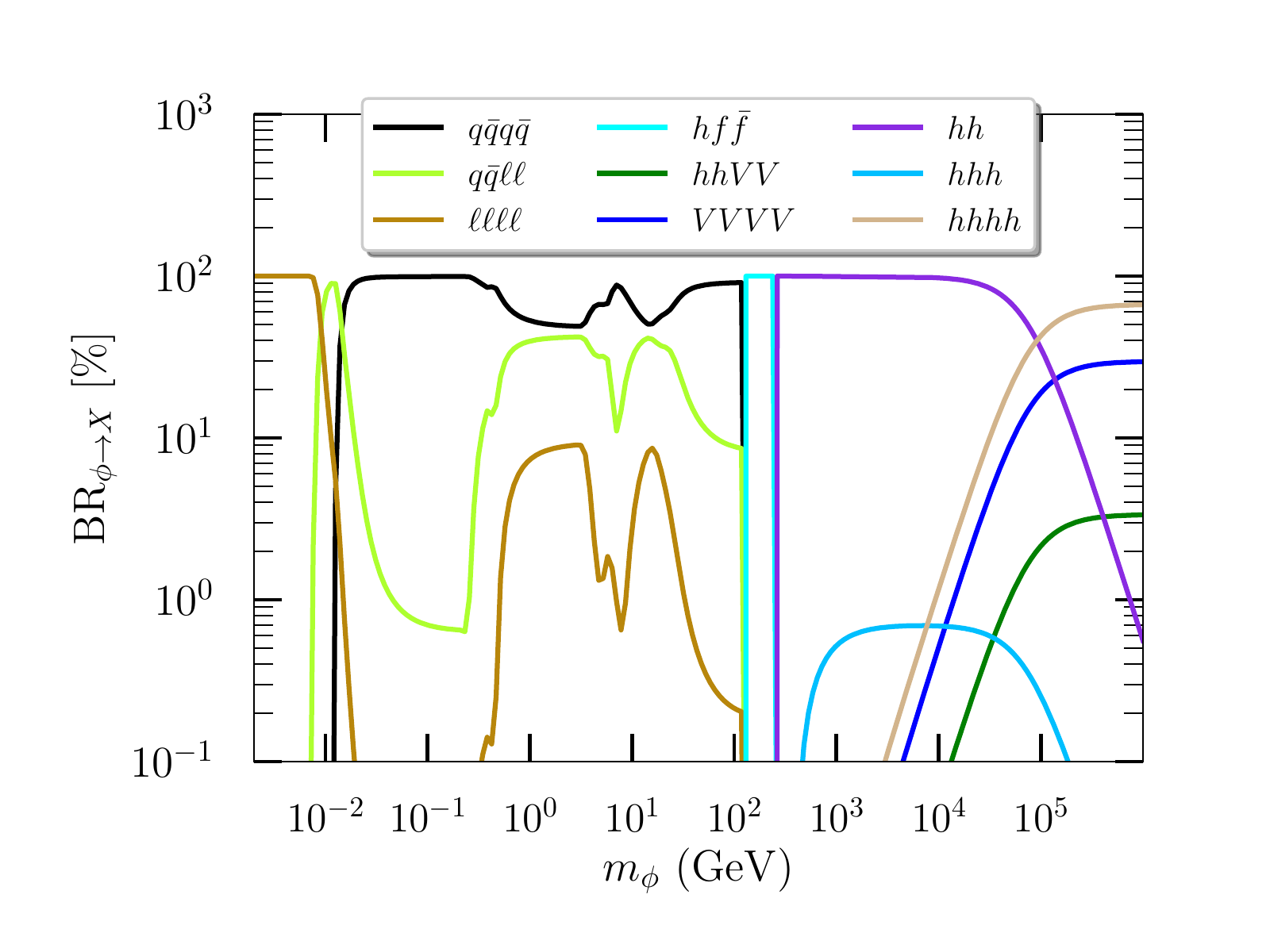}
\caption{Decay branching ratios of $\phi$ as function of $m_\phi$. Here, $f$ refers to $q, \ell$ and $V$ refers to $W^\pm/Z$. All these decays proceed through one or two Higgs bosons, i.e. $\phi \to q\bar{q} q\bar{q} \equiv \phi \to h^* (\to q\bar{q}) h^* (\to q\bar{q})$ (see the text for more details).}
\label{fig:DM_BR}
\end{figure}

The decay channels can be categorised as: $\phi \to hh$, $\phi \to h h^{*} \to h f\bar{f}, f VV, hhh$, and $\phi \to h^* h^* \to VVVV, VVf\bar{f}, VV hh, f\bar{f}f'\bar{f}', f\bar{f} hh, hhhh$, where $h$ being the SM Higgs boson, $V=W^\pm/Z$, and $f$ for SM-fermions, and the decay rates are suppressed by the factor $\xi^2/M^2$. However, besides the Higgs boson, DM does not decay \textit{directly} to all SM particles as in the case of metric gravity when performing Weyl transformaton \cite{ibarra}. 
As illustrated in Fig.\ref{fig:DM_BR}, one distinguishes three relevant DM mass ranges associated to the allowed decays:
\begin{itemize}
    \item $m_\phi \sim 10^{-3}-125$ GeV, in which the decay of DM is dominated by 4 light fermions.
    \item $m_\phi \sim 125-250$ GeV, where the DM decays predominantly into an on-shell Higgs boson and pair of light fermions.
    \item $m_\phi \sim 250-10^6$ GeV, where there are three competing decay channels $\phi \to hh$ with $\textrm{BR} \sim 100\%$ for $m_\phi \leq 10^4$ GeV, and $\phi \to 4 h$ ($\phi\to 4V$) whose branching fractions can reach up to $70\%$~($30\%$) for $m_\phi \sim 10^6$ GeV.
\end{itemize}

\begin{figure}[!t]
\centering
\includegraphics[width=\columnwidth]{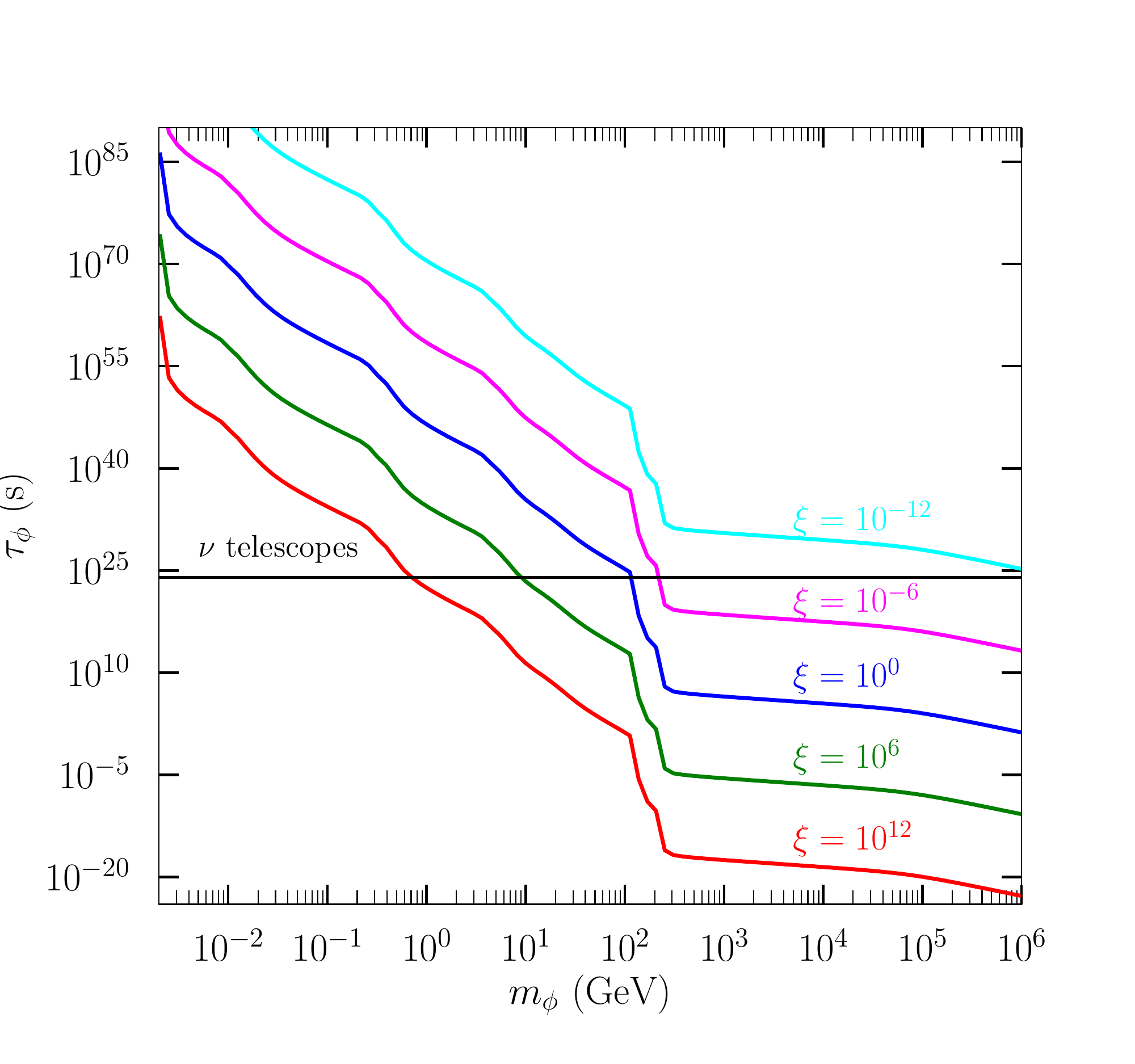}
\caption{DM lifetime $\tau_\phi$ computed from the inverse of the total decay width as a function of the DM mass $m_\phi$ for different values of $\xi$, with $M=M_{Pl}$. The horizontal black line shows the conservative upper bound on the DM lifetime from $\nu$-telescope experiments \cite{Esmaili:2012us}.}
\label{fig:DM_lifetime}
\end{figure}

The tiny decay rates resulting from the factor $M_{Pl}^{-2}$ implies a large lifetime for DM. There are several lower bounds on the DM lifetime, denoted here by $\tau_\phi$, from Fermi-LAT and AMS02 data \cite{Cirelli:2012ut, Ibarra:2013zia, Giesen:2015ufa}. These searches imply a mass-dependent bound on $\tau_\phi$ which cannot be  applied to our scenario since all the decay channels of DM generate, at least, four particles in the final state. A more comprehensive bounds require a systematic analysis of various astrophysical data while taking into account all the possible decay chains of the intermediate resonances (such as $W/Z/h$-bosons, and top quarks), parton showering, hadronisation and hadron decays, in addition to the possible uncertainties which may affect those rates \cite{Auchettl:2012sqm, Amoroso:2018qga}. This study is beyond the scope of this paper and will be done in a future work. We will take, however, a very conservative bound obtained from $\nu$-telescopes ($\tau > 10^{24}$ seconds). The dominance of the fermionic decay channels through the Higgs boson has very important consequences in our scenario especially for $m_\phi < 10$ GeV. The rationale for this is that the decay rates of DM get a suppression of order $y_f^4$ (with $y_f$ being the Yukawa coupling of the fermions to the SM-Higgs boson). This implies that the constraint on the parameter $\xi$ is very relaxed. In the heavy mass region, $m_\phi \sim 250-10^6$ GeV, the width becomes extremely large and therefore, one requires a very small coupling $\xi \leqslant 10^{-12}$, in order to be consistent with the existing bound. This finding can be clearly observed in Fig. \ref{fig:DM_lifetime} where $\tau_\phi$ is depicted in terms of the DM mass. We note that the scenario of light DM and relatively large couplings can be tested in current and future colliders in processes involving SM Higgs bosons such as $pp \to h^{*}(\to \phi h) Z/W^\pm$. 

\emph{Separable dark matter and standard model sectors}\textemdash   In this last part of the letter, we present briefly a general theoretical framework where both SM and DM sectors take place in the gravitational dynamics in a complete picture. Given a separable Lagrangians of the SM (coupled to metric) and DM (coupled to connection), one is able to construct a novel gravitational action principle including both sectors. In action (\ref{sm action}) the spacetime curvature is merely considered to be purely metric, however, in general this constraint must be lifted and one ends up with only one spacetime curvature where its connection is not metrical from scratch. Herein, the total action describing the complete sector is written as     
\begin{eqnarray}
S&&=\int d^{4}x
\label{total action}
\sqrt{|g|}\,\Big\{
\frac{M^2_{Pl}}{2}g^{\mu\nu}R_{\mu\nu}(\Gamma) +L[g,\text{SM}] \Big\} \nonumber \\ 
&&+ \int d^{4}x\, \mathcal{L}[\Gamma,\text{DM}],
\end{eqnarray}
where the Lagrangian densities of the SM and DM are given by their expressions (\ref{SM_lag}) and (\ref{DM_lag}) respectively. 

Consequently, variation with respect to metric leads to the gravitational field equations in which stress-energy is sourced only by the SM fields
\begin{eqnarray}
\label{einstein-palatini}
M^{2}_{Pl}R_{\mu\nu}(\Gamma)=
T_{\mu\nu}^{(\text{SM})}-\frac{1}{2}g_{\mu\nu}g^{\alpha\beta}T_{\alpha\beta}^{(\text{SM})}
\end{eqnarray}
whilst the DM appears via the dynamics of the affine connection arising from variation with respect to the connection itself 
\begin{flalign}
\label{dynamical equation}
\nabla_{\lambda}\left(\sqrt{|g|}\,g^{\mu\nu} +\frac{M^{2}}{M^{2}_{Pl}} \frac{\sqrt{\left| K\left(\Gamma,\phi\right) \right|}}{U(\phi)}
\left( K^{-1}\right)^{\mu\nu} \right)=0
\end{flalign}
where
\begin{eqnarray}
\label{tensor k}
K_{\mu\nu}\left(\Gamma,\phi\right)= M^{2} R_{\mu\nu}\left(\Gamma\right)- \nabla_{\mu}\phi\nabla_{\nu}\phi.
\end{eqnarray}

This means that the presence of DM action breaks the (metric) compatibility, and the solution of the last equation will bring the DM to the gravitational equations (\ref{einstein-palatini}) through the resulting Ricci curvature. Like in generalized Palatini theories of gravity, the presence of the Ricci curvature in the second term of (\ref{dynamical equation}) shows that this dynamical equation is not linear in the connection \cite{olmo}. Nevertheless, one can use the gravitational field equations (\ref{einstein-palatini}) to finally write the tensor field (\ref{tensor k}) in terms of matter only. Therefore, the previous equation becomes algebric equation that can be solved for the connection in terms of the matter sector. In fact, the connection is now the Levi-Civita of the tensor $\bm{h}$ (an auxiliary metric) written in a matrix form as  
\begin{eqnarray}
\hat{\bm{h}} = \left(\sqrt{\text{det}\hat{\bm{P}}} \right) \hat{\bm{P}}^{-1}\, \hat{g}, 
\end{eqnarray}
where the matrix $\hat{\bm{P}}$ includes the matter fields
\begin{eqnarray}
\hat{\bm{P}} = \hat{\bm{I}} + \frac{M^{2}}{M^{2}_{Pl} U(\phi)}\frac{\sqrt{|K|}}{\sqrt{|g|}}\, \hat{\bm{K}}.
\end{eqnarray}

Thus, the spacetime connection (coupled to DM) is solved explicitly in terms of the physical metric $g$ and matter sources described by both DM and SM fields. Interestingly, DM plays now a major role in determining the spacetime geometry which was not considered Riemannian \textit{a priori}.  Therefore, one can proceed with cosmological dynamics by considering different matter fields or cosmological fluids \cite{future:work}.

\emph{Conclusion}\textemdash  
%Understanding the way dark matter interacts to gravity may %reveal more interesting features of its physics. Indeed, its %pure gravitational effects remain still the only current %signature for its existence. 
In this letter, we have investigated the possibility where unlike ordinary matter, dark matter interacts to gravity through only the affine connection of spacetime, and decays to the standard model Higgs via a nonminimal coupling to gravity. Such interaction, which  is linear in the dark matter field and the Ricci curvature, and hence breaks the $Z_{2}$-symmetry, is generally permitted in curved spacetime.  The setup is fairly predictive in various levels:
(\textit{a}) dark matter can manifest simply as a scalar since scalars can fully decouple from metric but stay coupled to affine connection in the present framework, and  (\textit{b}) the induced decays occur  only through the SM  Higgs with tiny decay rates  especially for light dark matter implying relaxed constraints on the nonminimal coupling parameter as $\xi \leq 10^{12}$ for $m_\phi \leq 10$ GeV \textemdash which can be tested at current and future collider experiments, whereas for super-heavy dark matter (i.e. $m_\phi \geq 10$ TeV), it requires small values $\xi \leq 10^{-10}$.

The dark matter decay scenario we have presented here is fully gravitational and it stands on a different formulation of gravity. Indeed, it is not only metric, but spacetime connection plays also an important role in the gravitational couplings. It is thus assumed that dark matter may couple only to this metric-independent field. Last but not least, although we find that scalar singlets are the most viable candidate for dark matter in this context, the fact remains however that the present framework can be realised for femionic and vector dark matter when a complete theory of affine gravity assembling all matter fields is determined.

\section*{Acknowledgements}
HA is grateful to Durmu\c{s} Demir for useful discussion, and acknowledges support from the UAEU via startup grant No. 31S372. The work of AJ is supported by the National Research Foundation of Korea, Grant No. NRF-2019R1A2C1009419. 

%\newpage


\begin{thebibliography}{99}

\bibitem{planck}
N.~Aghanim \textit{et al.} [Planck],
\textit{Planck 2018 results. VI. Cosmological parameters},
[arXiv:1807.06209 [astro-ph.CO]].


\bibitem{rotation:curves} 
  V.~C.~Rubin, D.~Burstein, W.~K.~Ford, Jr. and N.~Thonnard,
  \textit{Rotation velocities of 16 SA galaxies and a comparison of Sa, Sb, and SC rotation properties},
  Astrophys. J. {\bf 289}, 81 (1985)
  

\bibitem{peebles} 
  P.~J.~E.~Peebles,
  \textit{Principles of physical cosmology},
  Princeton, USA: Univ. Pr. (1993)  

\bibitem{ibarra} O.~Catà, A.~Ibarra and S.~Ingenhütt,
  \textit{Dark matter decays from nonminimal coupling to gravity},
  Phys. Rev. Lett. {\bf 117}, no. 2, 021302 (2016);
  O.~Catà, A.~Ibarra and S.~Ingenhütt, \textit{Dark matter decay through gravity portals},
Phys. Rev. D \textbf{95}, no.3, 035011 (2017)
  
  \bibitem{demir_SM} 
  D.~A.~Demir, \textit{Effects of Curvature-Higgs Coupling on Electroweak Fine-Tuning},
  Phys. Lett. B {\bf 733}, 237 (2014).
  
  \bibitem{affine:inflation}
   H.~Azri and S.~Nasri, \textit{Entropy production in affine inflation},
  Phys.\ Rev.\ D {\bf 101}, no. 6, 064073 (2020);
  H.~Azri and D.~Demir,
  \textit{Induced Affine Inflation},
  Phys. Rev. D \textbf{97}, no.4, 044025 (2018);
  H.~Azri and D.~Demir,
  \textit{Affine Inflation},
  Phys.\ Rev.\ D {\bf 95}, no. 12, 124007 (2017);
  H. Azri, \textit{Are there really conformal frames? Uniqueness of affine inflation},
  Int.\ J.\ Mod.\ Phys.\ D {\bf 27}, no. 09, 1830006 (2018)
  
  \bibitem{fermions:without:matter} 
  D.~Amati and G.~Veneziano,
  \textit{Metric From Matter},
  Phys.\ Lett.\  {\bf 105B}, 358 (1981);
  J. Kijowski, \textit{On a new variational principle in general relativity and the energy of
gravitational field}, Gen.Rel.Grav.,9,857-877 (1978)

\bibitem{Cirelli:2012ut}
M.~Cirelli, E.~Moulin, P.~Panci, P.~D.~Serpico and A.~Viana,
\textit{Gamma ray constraints on Decaying Dark Matter},
Phys. Rev. D \textbf{86} (2012), 083506

\bibitem{Ibarra:2013zia}
A.~Ibarra, A.~S.~Lamperstorfer and J.~Silk,
\textit{Dark matter annihilations and decays after the AMS-02 positron measurements},
Phys. Rev. D \textbf{89} (2014) no.6, 063539

\bibitem{Giesen:2015ufa}
G.~Giesen, M.~Boudaud, Y.~Génolini, V.~Poulin, M.~Cirelli, P.~Salati and P.~D.~Serpico,
\textit{AMS-02 antiprotons, at last! Secondary astrophysical component and immediate implications for Dark Matter},
JCAP \textbf{09} (2015), 023

\bibitem{Auchettl:2012sqm}
K.~Auchettl and C.~Balázs,
\textit{Uncertainties in Dark Matter Indirect Detection},
doi:10.5772/52052

\bibitem{Amoroso:2018qga}
S.~Amoroso, S.~Caron, A.~Jueid, R.~Ruiz de Austri and P.~Skands,
\textit{Estimating QCD uncertainties in Monte Carlo event generators for gamma-ray dark matter searches},
JCAP \textbf{05} (2019), 007

\bibitem{Esmaili:2012us}
A.~Esmaili, A.~Ibarra and O.~Peres, L.G.,
\textit{Probing the stability of superheavy dark matter particles with high-energy neutrinos},
JCAP \textbf{11} (2012), 034

 
  \bibitem{olmo}
G.~J.~Olmo, H.~Sanchis-Alepuz and S.~Tripathi,
\textit{Dynamical Aspects of Generalized Palatini Theories of Gravity}, Phys. Rev. D \textbf{80}, 024013 (2009)
  
  \bibitem{future:work}
  Work in progress.



\end{thebibliography}
\end{document}